\documentclass[dvipsnames]{article}
\usepackage{adjustbox}
\usepackage{amssymb}
\usepackage{amsthm}
\usepackage{amsmath}
\usepackage{authblk}
\usepackage{blkarray}
\usepackage{booktabs}
\usepackage{bm}
\usepackage{calc}
\usepackage{caption}
\usepackage{cases}
\usepackage{centernot}
\usepackage[utf8]{inputenc}
\usepackage{dcolumn}
\usepackage{float}
\usepackage[margin=1in]{geometry}
\usepackage{graphicx}
\usepackage{listings}
\usepackage{makecell}
\usepackage{mathtools}
\usepackage{multicol}
\usepackage{pifont}

\usepackage{setspace}
\usepackage{soul}
\usepackage{subcaption}
\usepackage{tabularx}
\usepackage{tgpagella}
\usepackage{tikz}
    \usetikzlibrary{positioning}
\usepackage{tikzsymbols}
\usepackage{tkz-graph}
\usepackage{verbatim}
\usepackage{xcolor}
\usepackage{xfrac}
\usepackage{lineno}
\usepackage[style=english]{csquotes}
    \MakeOuterQuote{"}

\newcommand*\patchAmsMathEnvironmentForLineno[1]{%
  \expandafter\let\csname old#1\expandafter\endcsname\csname #1\endcsname
  \expandafter\let\csname oldend#1\expandafter\endcsname\csname end#1\endcsname
  \renewenvironment{#1}%
     {\linenomath\csname old#1\endcsname}%
     {\csname oldend#1\endcsname\endlinenomath}}%
\newcommand*\patchBothAmsMathEnvironmentsForLineno[1]{%
  \patchAmsMathEnvironmentForLineno{#1}%
  \patchAmsMathEnvironmentForLineno{#1*}}%
\AtBeginDocument{%
\patchBothAmsMathEnvironmentsForLineno{equation}%
\patchBothAmsMathEnvironmentsForLineno{align}%
\patchBothAmsMathEnvironmentsForLineno{flalign}%
\patchBothAmsMathEnvironmentsForLineno{alignat}%
\patchBothAmsMathEnvironmentsForLineno{gather}%
\patchBothAmsMathEnvironmentsForLineno{multline}%
}

\newcommand{\matr}[1]{\mathbf{#1}} 

\title{Convergence of reputations under indirect reciprocity}
\author[1,2]{Bryce Morsky \thanks{bmorsky@fsu.edu}}
\author[2]{Joshua B.\ Plotkin \thanks{jplotkin@sas.upenn.edu}}
\author[2]{Erol Ak\c{c}ay \thanks{eakcay@sas.upenn.edu}}
\affil[1]{Department of Mathematics, Florida State University, Tallahassee, FL, USA}
\affil[2]{Department of Biology, University of Pennsylvania, Philadelphia, PA, USA}
\date{\today}

\begin{document}

\maketitle

\begin{abstract}
    Previous research has shown how indirect reciprocity can promote cooperation through evolutionary game theoretic models. Most work in this field assumes a separation of time-scales: individuals' reputations equilibrate at a fast time scale for given frequencies of strategies while the strategies change slowly according to the replicator dynamics. Much of the previous research has focused on the behaviour and stability of equilibria for the replicator dynamics. Here we focus on the underlying reputational dynamics that occur on a fast time scale. We describe reputational dynamics as systems of differential equations and conduct stability analyses on their equilibria. We prove that reputations converge to a unique equilibrium for each of the five standard norms whether assessments are public or private. These results confirm a crucial but previously unconfirmed assumption underlying the theory of indirect reciprocity for the most studied set of norms.
\end{abstract}

{\textbf{Keywords:} cooperation, evolutionary game theory, indirect reciprocity, reputations, social norms}

\section{Introduction}

Indirect reciprocity is an important mechanism to foster cooperation. Theoretical studies have extended an evolutionary game theoretic model of the Prisoner's Dilemma game (also termed the Donation game) by adding a system of reputations and a population of Discriminators who defect against "bad" individuals and cooperate with "good" ones \cite{fishman03,leimar01,nowak05,ohtsuki06,okada20,okada18,sasaki17}. To determine who is good and who is bad, interactions between pairs of individuals are observed. To assess a donor's reputation (as good or bad), an observer may consider the action the donor took (to cooperate or defect), the reputation of the recipient with the observer (either good or bad), and a social norm. The social norms provide the rules of how to assess what the observer observed. For all norms, a donor is assessed as good if they cooperate with a good donor. This is a minimum requirement to promote cooperation. However, the norms will differ on their recommendations for other scenarios. There are five social norms that are frequently studied: Scoring \cite{wedekind00}, Shunning \cite{takahashi06}, Staying \cite{nakai08}, Simple Standing \cite{milinski01}, and Stern Judging \cite{santos16}. Scoring was the first norm studied, and under it a donor is considered good if they cooperate and bad if they defect. Assessments under Scoring do not depend on the reputation of the recipient, but this norm leads to a population that only ever defects. Thus, attention shifted to higher order norms that factor in the reputation of the recipient (Shunning, Staying, Simple Standing, and Stern Judging). Under the Stern Judging norm, donors are assessed as good if they cooperate with good recipients and defect with bad ones. Conversely, they are assessed as bad if they defect with good recipients and cooperate with bad ones. In the initial work studying these norms, reputations were generally assumed to be assessed publicly wherein there is a shared reputational system \cite{brandt04,chalub06,ohtsuki04} and all individuals agree on each other's reputation. Private reputations were later explored \cite{okada18}, which allow for individuals to hold private information and disagree about the reputations of others. Conflict between different opinions of individuals can undermine the reputational system and thus cooperation. These five norms and two methods of assessment (public and private) form the core set of theoretical models of indirect reciprocity. Additionally, there are a great many models extending this framework including noisy and incomplete information \cite{hilbe18}, individuals' emotions \cite{radzvilavicius19} and individuals' reasoning to account for errors and discrepancies in reputations \cite{morsky23,pandula24} to name a few. 

Despite a wide range of assumptions about assessment and game play, the above models generally share the assumption that the dynamics reputation and strategies operate at two different time scales. Reputations are assumed to equilibrate rapidly while strategies change much more slowly. This assumption can be justified if each individual undergoes many interactions during its lifetime (if replicator dynamics model a birth-death process, or simply update infrequently). A further justification is that individuals cannot fully assess the payoffs of their strategies (and thus not compare and imitate) until reputations have reached an equilibrium. As a theory building strategy, this assumption allows the models to account for the full incentive effects through reputations of a strategy composition of the population. It also makes the theory analytically tractable. 

Under this separation of time-scale assumption, the strategies of individuals change in response to payoffs that are computed for the frequencies of the types of individuals while assuming that their reputations are at the equilibrium levels given the strategy frequencies. Though equilibria of the reputational system were found previously and used to analyze the replicator dynamics that govern the change in strategies, whether or not they converged to these equilibria has been understudied. Convergence of reputations to unique equilibria has only been proved in a few cases such as for models that incorporate reasoning \cite{morsky23,pandula24}. However, convergence in some situations is conditional on the parameter values chosen \cite{pandula24}, and thus need not hold generally. Thus, it is an open and key question as to whether or not reputations converge. Here, we have proven that reputations in the standard indirect reciprocity model do indeed converge to unique equilibria for the five common norms and both public and private assessments of reputations. We do this by representing the reputational dynamics as a system of ordinary differential equations and analyse the stability of the equilibria of these systems. This setup places some assumptions on the reputation dynamics such as the population being infinite and time being continuous. However, these are frequently assumed in mathematical models of indirect reciprocity. So long as the population is large and dynamics occur in short intervals of time. In the methods section, we define the system of ordinary differential equations that model the reputational dynamics, and in the results section analyze the stability of their equilibria.

\section{Methods}

Consider three types of players each playing a specific strategy in the donation game. AllC (always cooperate) players are those who always \emph{intend} to cooperate regardless of the reputation of the recipient, and $x$ is their frequency in the population. Note that AllC players only intend to cooperate: they don't always successfully do so. As discussed below, we assume --- as is standard in the literature --- that errors may occur when players attempt to cooperate by which they unintentionally defect. AllD players always defect, and their frequency in the population is $y$. Note that there is no possibility for an AllD player to accidentally cooperate. The third and final type of player are Discriminators, who intend to cooperate with good recipients and defect against bad ones. They thus act as punishers of "bad behaviour" (as determined by the social norm). Their frequency in the population is $z$, and we have $x+y+z=1$. Since reputations converge rapidly, i.e.\ before strategies can change, $x$, $y$, and $z$ will be constants in our analyses.

Errors in action and assessment are assumed in many models of indirect reciprocity. Let $\tfrac{1}{2} > e_1 > 0$ be the probability that a donor who intends to cooperate defects by mistake. Further, let $\tfrac{1}{2} > e_2 > 0$ be the probability that there is an error in the assessment of the reputation of the donor. That is to say, with probability $e_2$, the observer assigns the \emph{opposite} reputation to the donor than they intended to. Define $\epsilon = (1-e_1)(1-e_2) + e_1e_2$ as the probability that an individual who intends to cooperate is observed doing so. We will use $\epsilon$ throughout our analysis rather than $e_1$. Also, we write $e=e_2$ to further simplify our notation. The parameters $\epsilon$ and $e$ along with the social norm and whether or not assessment are public will thus determine reputations.

\begin{table*}[!htpb]
\begin{center}
\begin{tabular}{lcccc}
&\multicolumn{4}{c}{\thead{Donor's action / recipient's reputation}} \\
Social norm & $C/G$ & $D/G$ & $C/B$ & $D/B$\\
\midrule
Scoring & $G$ & $B$ & $G$ & $B$\\
Shunning & $G$ & $B$ & $B$ & $B$\\
Simple Standing & $G$ & $B$ & $G$ & $G$\\
Staying & $G$ & $B$ & --- & ---\\
Stern Judging & $G$ & $B$ & $B$ & $G$\\
\end{tabular}
\end{center}
\vspace{-2mm}
\caption{Assessments of the donor (either $G$ or $B$ for good or bad) given the donor's action (either $C$ or $D$ for cooperate or defect), recipient's reputation ($G$ or $B$), and the social norm. The dash under Staying implies that the reputation of the donor is not updated when they interact with a recipient with a bad reputation.} \label{table:norms}
\end{table*}

Social norms determine what actions are good and what bad given the reputation of the donor. Five important norms frequently studied in the literature are: Scoring, Shunning, Staying, Simple Standing, and Stern Judging. The rules for these norms are represented in Table \ref{table:norms}. For example, under Simple Standing, a donor is assessed as good if they cooperate with a good recipient, and bad if they do not. And, they're assessed as good when they interact with a bad recipient, regardless of whether or not they cooperate. Note that due to the error in assessment, it is possible for an observer to assess a donor who interacts with a bad recipient as bad. However, in models where players can factor in error rates and thereby reason about the intention of the donor, this cannot happen \cite{morsky23,pandula24}.

The state variables in our analyses of the reputational dynamics are the fraction of players of each type that are good. Thus, $g_x$ and $1-g_x$ are the frequencies of AllC players with good and bad reputations, respectively. Likewise $g_y$ and $1-g_y$ are the frequencies of AllD players with good and bad reputations, respectively. Finally, $g_z$ and $1-g_z$ are the frequencies of Discriminators with good and bad reputations, respectively. The total frequency of good players in the population is $g=xg_x+yg_y+zg_z$.

Reputations can be assessed publicly or privately, which impact the assessment of the reputations of Discriminators because Discriminators' behaviours are determined by the reputation the recipient has with them. Thus, if this differs between the observer and a donor Discriminator, we need to know the frequency with which two players agree that a recipient is good, denoted $g_2$. The probability that two AllC players are good is defined as $g_{x2}$. $g_{y2}$ and $g_{z2}$ for AllD players and Discriminators are defined similarly. Thus, $g_2 = xg_{x2}+yg_{y2}+zg_{z2}$. Under private assessment of reputations, $g_{x2}$, $g_{y2}$, and $g_{z2}$ will be state variables in addition to $g_x$, $g_y$, and $g_z$.

As in previous models, we assume that reputations change by a good individual being reassessed as bad or a bad individual being reassessed as good \cite{okada18,sasaki17}. Let $g_i^+$ be the probability that a bad individual of type $i$ is reassessed as good, and $g_i^-$ be the probability that a good individual of type $i$ is reassessed as good. To represent this process as a system of differential equations, we define $\dot{g}_i = g_i^+-g_i^-$ by assuming a limiting process. We are thus able to convert a discrete process into a continuous one. We use the same values of $g_i^+$ and $g_i^-$ from \cite{okada18,sasaki17}. The reputational systems under public assessment for the five norms are thus:

\begin{subequations}
\noindent\begin{tabular}{@{}m{0.4\linewidth}@{}m{0.6\linewidth}@{}}
    \begin{equation}
    \begin{rcases}
        \dot{g}_x = \epsilon - g_x \\
        \dot{g}_y = e - g_y \\
        \dot{g}_z = \epsilon g + e(1-g) - g_z
    \end{rcases} \text{Scoring},
    \end{equation}
    &
    \begin{equation}
    \begin{rcases}
        \dot{g}_x = \epsilon g + e(1-g) - g_x \\
        \dot{g}_y = e - g_y \\
        \dot{g}_z = \epsilon g + e(1-g) - g_z
    \end{rcases} \text{Shunning},
    \end{equation}\\
    \begin{equation}
    \begin{rcases}
        \dot{g}_x = (\epsilon-g_x)g \\
        \dot{g}_y = (e-g_y)g \\
        \dot{g}_z = (\epsilon-g_z)g
    \end{rcases} \text{Staying},
    \end{equation}
    &
    \begin{equation}
    \begin{rcases}
        \dot{g}_x = \epsilon g + (1-e)(1-g) - g_x \\
        \dot{g}_y = eg + (1-e)(1-g) - g_y \\
        \dot{g}_z = \epsilon g + (1-e)(1-g) - g_z
    \end{rcases} \text{Simple Standing},
    \end{equation}
    \end{tabular}\\
    \noindent\begin{tabular}{@{}m{\linewidth}@{}}
    \begin{equation}
    \begin{rcases}
        \dot{g}_x = \epsilon g + (1-\epsilon)(1-g) - g_x \\
        \dot{g}_y = e g + (1-e)(1-g) - g_y \\
        \dot{g}_z = \epsilon g + (1-e)(1-g) - g_z
    \end{rcases} \text{Stern Judging}.
    \end{equation}
  \end{tabular}\\
\end{subequations}
And for private assessment of reputations, the systems are as follows:

\begin{subequations}
\noindent\begin{tabular}{@{}*{2}{m{0.5\linewidth}@{}}}
    \begin{equation}
    \begin{rcases}
        \dot{g}_x = \epsilon g + e(1-g) - g_x \\
        \dot{g}_y = e - g_y \\
        \dot{g}_z = \epsilon g_2 + e(1-g_2) - g_z \\
        \dot{g}_{x2} = (\epsilon g + e(1-g))g_x - g_{x2} \\
        \dot{g}_{y2} = eg_y - g_{y2} \\
        \dot{g}_{z2} = (\epsilon g_2 + e(1-g_2))g_z - g_{z2}
    \end{rcases} \text{Shunning},
    \end{equation}
    &
    \begin{equation}
    \begin{rcases}
        \dot{g}_x = (\epsilon-g_x)g \\
        \dot{g}_y = (e-g_y)g \\
        \dot{g}_z = \epsilon g_2 + e(g-g_2) - g_zg \\
        \dot{g}_{x2} = (\epsilon g_x - g_{x2})g \\
        \dot{g}_{y2} = (eg_y - g_{y2})g \\
        \dot{g}_{z2} = (\epsilon g_2 + e(g-g_2))g_z - g_{z2}g
    \end{rcases} \text{Staying},
    \end{equation}
    \end{tabular}\\
    \noindent\begin{tabular}{@{}m{\linewidth}@{}}
    \begin{equation}
    \begin{rcases}
        \dot{g}_x = \epsilon g + (1-e)(1-g) - g_x \\
        \dot{g}_y = eg + (1-e)(1-g) - g_y \\
        \dot{g}_z = \epsilon g_2 + e(g-g_2) + (1-e)(1-g) - g_z \\
        \dot{g}_{x2} = (\epsilon g + (1-e)(1-g))g_x - g_{x2} \\
        \dot{g}_{y2} = (eg + (1-e)(1-g))g_y - g_{y2} \\
        \dot{g}_{z2} = (\epsilon g_2 + e(g-g_2) + (1-e)(1-g))g_z - g_{z2}
    \end{rcases} \text{Simple Standing},
    \end{equation}
    \\
    \begin{equation}
    \begin{rcases}
        \dot{g}_x = \epsilon g + (1-\epsilon)(1-g) - g_x \\
        \dot{g}_y = eg + (1-e)(1-g) - g_y \\
        \dot{g}_z = (1-2g)(1-e) + g + (\epsilon-e)(2g_2-g) - g_z \\
        \dot{g}_{x2} = (\epsilon g + (1-\epsilon)(1-g))g_x - g_{x2} \\
        \dot{g}_{y2} = (eg + (1-e)(1-g))g_y - g_{y2} \\
        \dot{g}_{z2} = ((1-2g)(1-e) + g + (\epsilon-e)(2g_2-g))g_z - g_{z2}
    \end{rcases} \text{Stern Judging}.
    \end{equation}
  \end{tabular}\\
\end{subequations}
Since the reputation of the recipient is irrelevant for Scoring, the reputation dynamics are the same whether assessments are public or private.

To be more explicit on how these equations are derived, consider the dynamics for AllC under Simple Standing. $g_x$ increases when a bad AllC player is reassessed as good, which occurs with probability $g_x^+ = (1-g_x)(\epsilon g + (1-e)(1-g))$. A bad AllC player is selected with probability $1-g_x$. With probabilities $g$ and $1-g$ they pair with a good and bad recipient, respectively. When paired with a good recipient, the bad AllC player is reassessed as good with probability $\epsilon$. When paired with a good AllC player, they are reassessed as good with probability $1-e$. In a similarly fashion a good AllC player is reassessed as bad with probability $g_x^- = g_x((1-\epsilon)g + e(1-g))$. Simplifying we have $g_x^+ - g_x^- = \epsilon g + (1-e)g - g_x$, to which we define $\dot{g}_x$. The reputational dynamics for $g_y$ can be found similarly. Under public assessment of reputations, $g_z$ is also found similarly. However, under private assessment, one has to take care of how the Discriminator donor and observer view the reputations of the recipient. If they both agree that they're good, then the donor will intend to cooperate and the observer will evaluate them as if they're interacting with a good recipient. This occurs with probability $g_2$ and the donor will then be assessed as good with probability $\epsilon$. With probability $g-g_2$ the donor believes that the recipient is bad and so defects, but the observer believes that they're good. A donor who intends to defect against a good recipient will be assessed as good only if an error in assessment occurs, i.e.\ with probability $e$. Finally, with probability $1-g$, the observer believes that the recipient is bad, and thus will assess the donor as good so long as there is no error in assessment, i.e.\ with probability $1-e$. Thus, a bad Discriminator will be reassessed as good with probability $g_z^+ = (1-g_z)(\epsilon g_2 + e(g-g_2) + (1-e)(1-g))$. $g_z^-$ is found in a similar way. However, we consider the probabilities that the donor is bad. For example, if both donor and observer believe that the recipient is good, which happens with probability $g_2$, then the donor is assessed as bad with probability $1-\epsilon$. The other terms are found in a similar way giving us $g_z^- = g_z((1-\epsilon)g_2 + (1-e)(g-g_2) + e(1-g))$ and $\dot{g}_z = g_z^+-g_z^- = \epsilon g_2 + e(g_2-g) + (1-e)(1-g)$.

Continuing with the example of Simple Standing under private assessment of reputations, $g_x-g_{x2}$ is the probability that one player believes that an AllC player is good and another believes that they are bad. Thus, $g_{x2}$ increases when such an AllC player is reassessed as good, which happens with probability $g_{x2}^+ = (g_x-g_{x2})(\epsilon g + (1-e)(1-g))$. In a similar way, we can calculate the probability that $g_{x2}$ decreases as $g_{x2}^- = g_{x2}((1-\epsilon)g + e(1-g))$, which gives us $\dot{g}_{x2} = g_{x2}^+ - g_{x2}^- = (\epsilon g + (1-e)(1-g))g_x-g_{x2}$. $g_{y2}^+$, $g_{y2}^-$, $\dot{g}_{y2}$, $g_{z2}^+$, $g_{z2}^-$, and $\dot{g}_{z2}$ are all computed similarly.

\section{Results}

\subsection{Public Assessment}

First consider an analysis of the case of public assessment of reputations. Equilibria have previously been found in the literature \cite{sasaki17}, but here we show that reputations converge to a unique equilibrium for each norm. Beginning with Scoring, there is only one equilibrium frequency of good individuals, namely
\begin{equation}
    g^* = \frac{\epsilon x + e(1-x)}{1-(\epsilon-e)z},
\end{equation}
which is defined across the simplex and for all error rates (as will all equilibria here). The Jacobian of the system of ODEs is
\begin{equation}
    \matr{J} = \begin{pmatrix}
    -1 & 0 & 0 \\
    0 & -1 & 0 \\
    (\epsilon-e)x & (\epsilon-e)y & (\epsilon-e)z - 1 \\
    \end{pmatrix},
\end{equation}
which has eigenvalues $\lambda_1 = \lambda_2 = -1$ and $\lambda_3 = (\epsilon-e)z-1 < 0$. Note that $\matr{J}$ is not a function of $g$, which will be the case under public assessment for all norms but Staying. Since the eigenvalues are negative and $g^*$ is the sole equilibrium, $g^*$ is stable. Additionally, since assessments under Scoring do not depend upon the reputation of the recipient, there is no difference in Scoring between public and private assessments and thus these results hold for both. 

For Shunning, the sole equilibrium frequency of good individuals is
\begin{equation}
    g^* = \frac{e}{1-(\epsilon-e)(1-y)},
\end{equation}
and the Jacobian of the system is
\begin{equation}
    \matr{J} = \begin{pmatrix}
    (\epsilon-e)x - 1 & (\epsilon -e)y & (\epsilon -e)z \\
    0 & -1 & 0 \\
    (\epsilon -e)x & (\epsilon -e)y & (\epsilon-e)z -1 \\
    \end{pmatrix}.
\end{equation}
It has eigenvalues $\lambda_1 = \lambda_2 = -1$ and $\lambda_3 = (\epsilon-e)(1-y) - 1 < 0$, and thus $g^*$ is stable.

For Staying, there are two equilibria: $g^* = 0$ and $g^* = \epsilon(1-y) + ey$, and in the latter case $g_x^*=g_z^*=\epsilon$ and $g_y^*=e$. Subbing these solutions into the Jacobian gives us the following matrices:
\begin{subequations}
\begin{equation}
    \matr{J}(0) = \begin{pmatrix}
    \epsilon x & \epsilon y & \epsilon z \\
    e x & e y & e z \\
    \epsilon x & \epsilon y & \epsilon z \\
    \end{pmatrix}, \label{Pub_Sh_J0}
    \end{equation}
    \begin{equation}
    \matr{J}(\epsilon(1-y) + ey) = \begin{pmatrix}
    -\epsilon(1-y) - e y & 0 & 0 \\
    0 & -\epsilon(1-y) - e y & 0 \\
    0 & 0 & -\epsilon(1-y) - e y \\
    \end{pmatrix}. \label{Pub_Sh_Jint}
    \end{equation}
\end{subequations}
The eigenvalues for Equation \ref{Pub_Sh_J0} are $\lambda_1 = \lambda_2 = 0$ and $\lambda_3 = \epsilon(1-y) + ey > 0$, and thus $g^*=0$ is unstable. The eigenvalues for Equation \ref{Pub_Sh_Jint} are $\lambda_1 = \lambda_2 = \lambda_3 = - (1-y)\epsilon - ey < 0$, and thus $g^*=\epsilon(1-y)+ey$ is stable. 

For Simple Standing, we have the sole equilibrium
\begin{equation}
    g^* = \frac{1-e}{1 - e + 1 - \epsilon(1-y) - ey}.
\end{equation}
The Jacobian of the system is
\begin{equation}
    \matr{J} = \begin{pmatrix}
    (\epsilon+e-1)x - 1 & (\epsilon+e-1)y & (\epsilon+e-1)z \\
    (2e-1)x & (2e-1)y - 1 & (2e-1)z \\
    (\epsilon+e-1)x & (\epsilon+e-1)y & (\epsilon+e-1)z - 1 \\
    \end{pmatrix},
\end{equation}
which has eigenvalues $\lambda_1 = \lambda_2 = -1$ and $\lambda_3 = e - 1 + \epsilon(1-y) + e y - 1 < 0$. Therefore, $g^*$ is stable.

And, finally, for Stern Judging, we have the sole equilibrium
\begin{equation}
    g^* = \frac{1 - \epsilon x - e(1-x)}{1 - \epsilon x - e(1-x) + 1 - \epsilon(1-y) - e y}.
\end{equation}
The Jacobian of the system is
\begin{equation}
    \matr{J} = \begin{pmatrix}
    (2\epsilon-1)x - 1 & (2\epsilon-1)y & (2\epsilon-1)z \\
    (2e-1)x & (2e-1)y - 1 & (2e-1)z \\
    (\epsilon+e-1)x & (\epsilon+e-1)y & (\epsilon+e-1)z - 1 \\
    \end{pmatrix},
\end{equation}
which has eigenvalues $\lambda_1 = \lambda_2 = -1$ and $\lambda_3 = \epsilon x + e(1-x) - 1 + \epsilon(1-y) + ey - 1 < 0$. Therefore, $g^*$ is stable.

\subsection{Private Assessment}

For Shunning, the sole equilibrium frequency of good individuals is
\begin{equation}
    g^* = \frac{\epsilon g_2^*z + e(1-g_2^*z)}{1-(\epsilon-e)x}.
\end{equation}
The Jacobian of the system evaluated at this equilibrium is
\begin{equation}
    \matr{J}(g^*) = \begin{pmatrix}
    (\epsilon-e)x-1 & (\epsilon-e)y & (\epsilon-e)z & 0 & 0 & 0 \\
    0 & -1 & 0 & 0 & 0 & 0 \\
    0 & 0 & -1 & (\epsilon-e)x & (\epsilon-e)y & (\epsilon-e)z \\
    ((\epsilon-e)x+1)g_x^* & (\epsilon-e)g_x^*y & (\epsilon-e)g_x^*z & -1 & 0 & 0 \\
    0 & e & 0 & 0 & -1 & 0 \\
    0 & 0 & g_z^* & (\epsilon-e)g_z^*x & (\epsilon-e)g_z^*y & (\epsilon-e)g_z^*z-1 \\
    \end{pmatrix}.
\end{equation}
The characteristic equation is $(1+\lambda)^3(\lambda^3 + c_2\lambda^2 + c_1\lambda + c_0)=0$ with the following coefficients:
\begin{subequations}
\begin{align}
    c_2 &= 3-(\epsilon-e)(x+g_z^*z) > 2, \\
    c_1 &= 3 - (\epsilon-e)(3g_z^*z + (2+(\epsilon-e)(g_x^*-g_z^*)z)x) > 0, \\
    c_0 &= 1 - (\epsilon-e)(x + 2g_z^*z + 2(\epsilon-e)(g_x^*-g_z^*)xz) \notag \\
    &= 1 - (\epsilon-e)(2ez+x(1+2g^*z(\epsilon-e)^2)) - 2z(\epsilon-e)^2(1 - (\epsilon-e)x)g_2^* \notag \\
    &\geq 1 - (\epsilon-e)(2ez+x(1+2g^*z(\epsilon-e)^2)) - 2z(\epsilon-e)^2(1 - (\epsilon-e)x)g^* \notag \\
    &= 1 - (\epsilon-e)x - 2(\epsilon-e)((\epsilon-e)g^*+e)z \notag \\
    &\geq 1 - (\epsilon-e)x - \frac{2e(\epsilon-e)}{1-\epsilon+e}z \notag \\
    &> 1 - (\epsilon-e)x - z \geq 0. 
\end{align}
\end{subequations}
The inequality for $c_0$ follows for the following reasons: $g_x^*=(\epsilon-e)g^*+e$ and $g_z^*=(\epsilon-e)g_2^*+e$; $g^* \geq g_2^*$; $g_x^* = (\epsilon-e)g^* + e \leq (\epsilon-e)g_x^* + e \implies g_x^* \leq e/(1-\epsilon+e)$; and $1-\epsilon+e - 2e(\epsilon-e) = e_1 + 4(1-e_1)e^2 > 0$. The first three eigenvalues are negative as can be seen from the first factor of the characteristic equation. The second factor is a cubic equation of $\lambda$. Note that all of the coefficients of this cubic are positive. The Routh-Hurwitz criterion for stability requires that all coefficients of this cubic to be positive and $c_2c_1-c_0 > 0$. Checking this last condition gives us
\begin{equation}
    c_2c_1-c_0 > 2c_1-c_0 = 5 - (\epsilon-e)(3x + 4g_z^*z) > 0.
\end{equation}
Therefore, $g^*$ is stable.

There are two equilibria for Staying under private assessment. The first of which is $g^*=0$, and evaluating the Jacobian at this equilibrium gives us
\begin{equation}
    \matr{J}(0) = \begin{pmatrix}
    x \epsilon  & y \epsilon  & z \epsilon  & 0 & 0 & 0 \\
    e x & e y & e z & 0 & 0 & 0 \\
    e x & e y & e z & (\epsilon-e)x & (\epsilon-e)y & (\epsilon-e)z \\
    0 & 0 & 0 & 0 & 0 & 0 \\
    0 & 0 & 0 & 0 & 0 & 0 \\
    0 & 0 & 0 & 0 & 0 & 0 \\
    \end{pmatrix}.
\end{equation}
The eigenvalues are $\lambda_i = 0$ for $i=1\ldots,5$ and $\lambda_6 = \epsilon x + e(1-x) > 0$. Thus, $g^*=0$ is unstable. At the other equilibrium, $g_x^*=\epsilon$, $g_y^*=e$, $g_z^*=(\epsilon-e)g_2^*/g^* + e$, and the Jacobian evaluated at it is
\begin{subequations}
\begin{align}
    \matr{J}(g^*) &= \begin{pmatrix}
        \matr{J}_1 & \matr{J}_2 \\
        \matr{J}_3 & \matr{J}_4
    \end{pmatrix}, \\
    \matr{J}_1 &= \begin{pmatrix}
    -g^* & 0 & 0 \\
    0 & -g^* & 0 \\
    (e-g_z^*)x & (e-g_z^*)y & (e-g_z^*)z-g^* \\
    \end{pmatrix}, \\
    \matr{J}_2 &= \begin{pmatrix}
    0 & 0 & 0 \\
    0 & 0 & 0 \\
    (\epsilon-e)x & (\epsilon-e)y & (\epsilon-e)z \\
    \end{pmatrix}, \\
    \matr{J}_3 &= \begin{pmatrix}
    \epsilon g^* & 0 & 0 \\
    0 & eg^* & 0 \\
    (e-g_z^*)g_z^*x & (e-g_z^*)g_z^*y & (e-g_z^*)g_z^*z + g_z^*g^* \\
    \end{pmatrix}, \\
    \matr{J}_4 &= \begin{pmatrix}
    -g^* & 0 & 0 \\
    0 & -g^* & 0 \\
    (\epsilon-e)g_z^*x & (\epsilon-e)g_z^*y & (\epsilon-e)g_z^*z-g^* \\
    \end{pmatrix}.
\end{align}
\end{subequations}
The characteristic equation is $(\lambda+g^*)^4(\lambda^2 + c_1\lambda+c_0) = 0$. Since $g^* \geq g_y^* = e$,
\begin{subequations}
\begin{align}
    c_1 &= 2g^* - ez + (1-\epsilon+e)g_z^*z \geq e(2-z) + (1-\epsilon+e)g_z^*z > 0, \\
    c_0 &= g^*(g^* + ((1 - 2(\epsilon-e))g_z^* - e)z) \geq g^*(g_z^*z + ((1 - 2(\epsilon-e))g_z^* - e)z) = g^*\sqrt{k_1+k_2+k_3} > 0, \label{private_staying_c0} \\
    k_1 &= k1 = e^2(y-z)^2 + 2\epsilon e xy + \epsilon^2 x^2  > 0, \\
    k_2 &= 2e^2 yz(2(1-\epsilon^2) + 2e(2\epsilon-e)) > 0, \\
    k_3 &= 2\epsilon exz + 4\epsilon(1-\epsilon)(\epsilon-e)^2 xz > 0.
\end{align}
\end{subequations}
We obtain the inequality for Equation \ref{private_staying_c0} by solving $g^*=\epsilon x + ey + g_z^*z$ and $g_2^*=\epsilon^2x + e^2y + (g_z^*)^2z = \epsilon^2x + e^2y + ((\epsilon-e)g_2^*/g^* + e)^2z$ for $g^*$ and $g_2^*$ and then plugging these solutions into $c_0$. Note also that the radicand is positive. Thus, the last two eigenvalues must be negative and so $g^*$ is stable.

Next consider Simple Standing. The equilibrium frequency of good individuals is
\begin{equation}
    g^* = \frac{1-e + (\epsilon-e)zg_2^*}{1-2e+1-(\epsilon-e)},
\end{equation}
and the Jacobian evaluated at it is
\begin{subequations}
\begin{align}
    \matr{J}(g^*) &= \begin{pmatrix}
        \matr{J}_1 & \matr{J}_2 \\
        \matr{J}_3 & \matr{J}_4
    \end{pmatrix}, \\
    \matr{J}_1 &= \begin{pmatrix}
    (\epsilon+e-1)x-1 & (\epsilon+e-1)y & (\epsilon+e-1)z \\
    (2e-1)x & (2e-1)y-1 & (2e-1)z \\
    (2e-1)x & (2e-1)y & (2e-1)z-1 \\
    \end{pmatrix}, \\
    \matr{J}_2 &= \begin{pmatrix}
    0 & 0 & 0 \\
    0 & 0 & 0 \\
    (\epsilon-e)x & (\epsilon-e)y & (\epsilon-e)z \\
    \end{pmatrix}, \\
    \matr{J}_3 &= \begin{pmatrix}
    (\epsilon+e-1)g_x^*x+g_x^* & (\epsilon+e-1)g_x^*y & (\epsilon+e-1)g_x^*z \\
    (2e-1)g_y^*x & (2e-1)g_y^*y+g_y^* & (2e-1)g_y^*z \\
    (2e-1)g_z^*x & (2e-1)g_z^*y & (2e-1)g_z^*z+g_z^* \\
    \end{pmatrix}, \\
    \matr{J}_4 &= \begin{pmatrix}
    -1 & 0 & 0 \\
    0 & -1 & 0 \\
    (\epsilon-e)g_z^*x & (\epsilon-e)g_z^*y & (\epsilon-e)g_z^*z-1 \\
    \end{pmatrix}.
\end{align}
\end{subequations}
The characteristic equation is $(\lambda+1)^3(\lambda^3 + c_2\lambda^2+c_1\lambda+c_0) = 0$ with positive coefficients:
\begin{subequations}
\begin{align}
    c_2 &= 2 + 1 - (\epsilon-e)g_z^*z + (1-\epsilon-e)x + (1-2e)y + (1-2e)z > 2, \\
    c_1 &= (1+(1-\epsilon-e)(2+(\epsilon-e)(g_x^*-g_z^*)z))x + 2(1-2e)(1-(\epsilon-e)(g_z^*-g_y^*)z)y \notag \\
    &+ 3(1-(\epsilon-e)g_z^*z) + 2(1-2e)z >0, \\
    c_0 &= (1+(1-\epsilon-e)(1 + 2(\epsilon-e)(g_x^*-g_z^*)z))x + 2(e + (1-2e)(1-(\epsilon-e)(g_z^*-g_y^*)z))y \notag \\
    &+ 2(1 - \epsilon g_z^* - e(1-g_z^*))z > 0,
    \end{align}
\end{subequations}
since $1-\epsilon-e = e_1(1-2e) > 0$. The first three eigenvalues are negative. Since all of the coefficients of the cubic are positive, we need only to confirm that $c_2c_1-c_0 > 2c_1-c_0 > 0$ to prove stability. Checking this last condition gives us
\begin{align}
    2c_1-c_0 &= 8 - 6e - 3x(\epsilon-e) - 4(\epsilon-e)(1-x-y)g_z^* \notag \\
            &\geq 8 - 6e - 3x(\epsilon-e) - 4(\epsilon-e)(1-x-y) \notag \\
            &= 4-2e + 4(1-\epsilon) + (\epsilon-e)(x+4y) > 0.
\end{align}
Therefore, it is stable.

Finally, consider Stern Judging, which has the equilibrium $g^*=g_x^*=g_y^*=g_z^*=\tfrac{1}{2}$ \cite{okada18}. Evaluating the Jacobian at this equilibrium gives us
\begin{subequations}
\begin{align}
    \matr{J}(g^*) &= \begin{pmatrix}
        \matr{J}_1 & \matr{J}_2 \\
        \matr{J}_3 & \matr{J}_4
    \end{pmatrix}, \\
    \matr{J}_1 &= \begin{pmatrix}
    (2\epsilon-1)x-1 & (2\epsilon-1)y & (2\epsilon-1)z \\
    (2e-1)x & (2e-1)y-1 & (2e-1)z \\
    (2e-1+e-\epsilon)x & (2e-1+e-\epsilon)y & (2e-1+e-\epsilon)z-1 \\
    \end{pmatrix}, \\
    \matr{J}_2 &= \begin{pmatrix}
    0 & 0 & 0 \\
    0 & 0 & 0 \\
    2(\epsilon-e)x & 2(\epsilon-e)y & 2(\epsilon-e)z \\
    \end{pmatrix}, \\
    \matr{J}_3 &= \begin{pmatrix}
    (2\epsilon-1)g_x^*x+g_x^* & (2\epsilon-1)g_x^*y & (2\epsilon-1)g_x^*z \\
    (2e-1)g_y^*x & (2e-1)g_y^*y+g_y^* & (2e-1)g_y^*z \\
    (2e-1+e-\epsilon)g_z^*x & (2e-1+e-\epsilon)g_z^*y & (2e-1+e-\epsilon)g_z^*z+g_z^* \\
    \end{pmatrix}, \\
    \matr{J}_4 &= \begin{pmatrix}
    -1 & 0 & 0 \\
    0 & -1 & 0 \\
    2(\epsilon-e)g_z^*x & 2(\epsilon-e)g_z^*y & 2(\epsilon-e)g_z^*z-1 \\
    \end{pmatrix}.
\end{align}
\end{subequations}
The characteristic equation is $(\lambda+1)^2(\lambda^4 + c_3\lambda^3 + c_2\lambda^2 + c_1\lambda + c_0)=0$ with coefficients:
\begin{subequations}
\begin{align}
    c_3 &= 4-\epsilon x + (1-\epsilon)x + (1-2e)(1-x) > 0, \\
    c_2 &= 3 + 3(1-\epsilon x) + 3(1-\epsilon)x + 3(1-2e)y + (2(1-2e)+1-\epsilon-e)z > 3, \\
    c_1 &= 3(1-\epsilon x) + 3(1-\epsilon)x + 3(1-2e)y + 1-\epsilon z + (2(1-2e)+1-\epsilon)z > 0 \\
    c_0 &= 1-\epsilon x +(1-\epsilon)x + (1-2e)y + (1-\epsilon-e)z > 0,
    \end{align}
\end{subequations}
Further, we have the following inequalities:
\begin{subequations}
\begin{align}
    &c_3c_2 - c_1 \geq 3c_3 - c_1 = 8 - 2(\epsilon-e)z > 0, \\
    &c_3c_2c_1 - c_3^2c_0 - c_1^2 = k_1k_2 > 0, \\
    &k_1 = 2(2 - (2\epsilon-1)x + (1-2e)(y+z))  > 0, \\
    &k_2 = (4 + 2(1-2\epsilon)x + 2(1-2e)y + (1-2e + 1-\epsilon-e)z))^2 > 0.
\end{align}
\end{subequations}
Therefore, $g^*$ is stable by the Routh-Hurwitz criteria.

\section{Discussion}

Indirect reciprocity is a key mechanism to promote cooperation and has been well studied in the literature with both theoretical models and experiments. Models have shown how indirect reciprocity can evolve and how it can promote cooperation. Additionally, experimental evidence of indirect reciprocity has been found in both humans and other animals \cite{akcay10,nava19,seinen06,sommerfeld07,yoeli13}. Many of the mathematical models of indirect reciprocity assume a fast dynamic for reputations and a slow dynamic for strategies. That is to say, reputations of individuals are assessed and reach an equilibrium relatively quickly. Expected payoffs are calculated given these reputations. Then, individuals can change their strategies by imitating those who have greater payoffs. Reputations converge to an equilibrium quickly again, and so on. These models assume that reputations converge to a unique equilibrium, but this was not proven for either public or private assessment of reputations. Here, we closed this gap, and have shown that the reputational dynamics that occur rapidly do converge to unique equilibria for each of the five standard norms and two assessment rules, which provides a basis for the previous analysis of the strategical dynamics in the literature.

Extensions to the models may be qualitatively different than the reputational dynamics explored here. For example, the norms we considered are zeroth order (Scoring) and first order (Shunning, Staying, Simple Standing, and Stern Judging). Assignment of reputations under Scoring only depends upon the action of the donor while assignments of the other norms also depends on the reputation of the recipient. Higher order norms --- those that use other information such as the previous reputation of the donor or multiple observations of the donor --- may not lead to convergence to a unique set of reputations. For third order norms, the reputational system can be bistable, and when observers make multiple observations, reputations may not converge (unpublished research). We note that that our systems of ODEs contain at most cubic polynomials with respect to the variables. The case of multiple observations is quartic, while the abductive reasoning model --- which has conditional convergence --- involves rational functions \cite{pandula24}. Finding higher order norms that are relevant to behaviour and that do not converge to reputational equilibria is an area for future research. Another research area that may lead to non-convergence is finite populations whether well-mixed or on a network. It's possible that small populations do not converge to unique equilibria or that there are network configurations that also impede convergence.

\subsection*{Code and data availability}
Code to verify analytical results is available at github.com/bmorsky/indirectReciprocity-convergence.

\bibliography{converge}

\begin{thebibliography}{10}

\bibitem{akcay10}
{\c{C}}a{\u{g}}lar Ak{\c{c}}ay, Veronica~A Reed, S~Elizabeth Campbell,
  Christopher~N Templeton, and Michael~D Beecher.
\newblock Indirect reciprocity: song sparrows distrust aggressive neighbours
  based on eavesdropping.
\newblock {\em Animal Behaviour}, 80(6):1041--1047, 2010.

\bibitem{brandt04}
Hannelore Brandt and Karl Sigmund.
\newblock The logic of reprobation: assessment and action rules for indirect
  reciprocation.
\newblock {\em Journal of theoretical biology}, 231(4):475--486, 2004.

\bibitem{chalub06}
Fabio~ACC Chalub, Francisco~C Santos, and Jorge~M Pacheco.
\newblock The evolution of norms.
\newblock {\em Journal of theoretical biology}, 241(2):233--240, 2006.

\bibitem{fishman03}
Michael~A Fishman.
\newblock Indirect reciprocity among imperfect individuals.
\newblock {\em Journal of Theoretical Biology}, 225(3):285--292, 2003.

\bibitem{hilbe18}
Christian Hilbe, Laura Schmid, Josef Tkadlec, Krishnendu Chatterjee, and
  Martin~A Nowak.
\newblock Indirect reciprocity with private, noisy, and incomplete information.
\newblock {\em Proceedings of the national academy of sciences},
  115(48):12241--12246, 2018.

\bibitem{leimar01}
Olof Leimar and Peter Hammerstein.
\newblock Evolution of cooperation through indirect reciprocity.
\newblock {\em Proceedings of the Royal Society of London. Series B: Biological
  Sciences}, 268(1468):745--753, 2001.

\bibitem{milinski01}
Manfred Milinski, Dirk Semmann, Theo~CM Bakker, and Hans-J{\"u}rgen Krambeck.
\newblock Cooperation through indirect reciprocity: image scoring or standing
  strategy?
\newblock {\em Proceedings of the Royal Society of London. Series B: Biological
  Sciences}, 268(1484):2495--2501, 2001.

\bibitem{morsky23}
Bryce Morsky, Joshua~B Plotkin, and Erol Akcay.
\newblock Indirect reciprocity with bayesian reasoning and biases.
\newblock {\em SocArXiv. October}, 16, 2023.

\bibitem{nakai08}
Yutaka Nakai and Masayoshi Muto.
\newblock Emergence and collapse of peace with friend selection strategies.
\newblock {\em Journal of Artificial Societies and Social Simulation}, 11(3):6,
  2008.

\bibitem{nava19}
Elena Nava, Emanuela Croci, and Chiara Turati.
\newblock ‘{I} see you sharing, thus {I} share with you’: indirect
  reciprocity in toddlers but not infants.
\newblock {\em Palgrave Communications}, 5(1):1--9, 2019.

\bibitem{nowak05}
Martin~A Nowak and Karl Sigmund.
\newblock Evolution of indirect reciprocity.
\newblock {\em Nature}, 437(7063):1291--1298, 2005.

\bibitem{ohtsuki04}
Hisashi Ohtsuki and Yoh Iwasa.
\newblock How should we define goodness?—reputation dynamics in indirect
  reciprocity.
\newblock {\em Journal of theoretical biology}, 231(1):107--120, 2004.

\bibitem{ohtsuki06}
Hisashi Ohtsuki and Yoh Iwasa.
\newblock The leading eight: social norms that can maintain cooperation by
  indirect reciprocity.
\newblock {\em Journal of theoretical biology}, 239(4):435--444, 2006.

\bibitem{okada20}
Isamu Okada.
\newblock A review of theoretical studies on indirect reciprocity.
\newblock {\em Games}, 11(3):27, 2020.

\bibitem{okada18}
Isamu Okada, Tatsuya Sasaki, and Yutaka Nakai.
\newblock A solution for private assessment in indirect reciprocity using
  solitary observation.
\newblock {\em Journal of theoretical biology}, 455:7--15, 2018.

\bibitem{pandula24}
Neel Pandula, Erol Ak{\c{c}}ay, and Bryce Morsky.
\newblock Indirect reciprocity with abductive reasoning.
\newblock {\em Journal of Theoretical Biology}, 580:111715, 2024.

\bibitem{radzvilavicius19}
Arunas~L Radzvilavicius, Alexander~J Stewart, and Joshua~B Plotkin.
\newblock Evolution of empathetic moral evaluation.
\newblock {\em eLife}, 8:e44269, 2019.

\bibitem{santos16}
Fernando~P Santos, Jorge~M Pacheco, and Francisco~C Santos.
\newblock Evolution of cooperation under indirect reciprocity and arbitrary
  exploration rates.
\newblock {\em Scientific reports}, 6(1):37517, 2016.

\bibitem{sasaki17}
Tatsuya Sasaki, Isamu Okada, and Yutaka Nakai.
\newblock The evolution of conditional moral assessment in indirect
  reciprocity.
\newblock {\em Scientific reports}, 7(1):1--8, 2017.

\bibitem{seinen06}
Ingrid Seinen and Arthur Schram.
\newblock Social status and group norms: Indirect reciprocity in a repeated
  helping experiment.
\newblock {\em European economic review}, 50(3):581--602, 2006.

\bibitem{sommerfeld07}
Ralf~D Sommerfeld, Hans-J{\"u}rgen Krambeck, Dirk Semmann, and Manfred
  Milinski.
\newblock Gossip as an alternative for direct observation in games of indirect
  reciprocity.
\newblock {\em Proceedings of the national academy of sciences},
  104(44):17435--17440, 2007.

\bibitem{takahashi06}
Nobuyuki Takahashi and Rie Mashima.
\newblock The importance of subjectivity in perceptual errors on the emergence
  of indirect reciprocity.
\newblock {\em Journal of Theoretical Biology}, 243(3):418--436, 2006.

\bibitem{wedekind00}
Claus Wedekind and Manfred Milinski.
\newblock Cooperation through image scoring in humans.
\newblock {\em Science}, 288(5467):850--852, 2000.

\bibitem{yoeli13}
Erez Yoeli, Moshe Hoffman, David~G Rand, and Martin~A Nowak.
\newblock Powering up with indirect reciprocity in a large-scale field
  experiment.
\newblock {\em Proceedings of the National Academy of Sciences}, 110(Supplement
  2):10424--10429, 2013.

\end{thebibliography}
\bibliographystyle{plain}

\end{document}